\documentclass[journal,onecolumn]{IEEEtran}
\usepackage[utf8]{inputenc}

\usepackage{amsmath}
\usepackage{amssymb}
\usepackage{bm}

\usepackage{amsthm}

\newtheorem{theorem}{Theorem}

\newtheorem{definition}{Definition}
\newtheorem{lemma}{Lemma}
\theoremstyle{remark}
\newtheorem{remark}{Remark}

\usepackage{algorithm}
\usepackage{algorithmic}

\usepackage{eqparbox}

\usepackage{color}
 \newcommand{\sd}{}
 
\usepackage{graphicx}
\usepackage{float}
\newcommand{\qf}[1]{{\color{blue}#1}}

\DeclareMathOperator*{\minimize}{minimize}

\title{Coded Alternating Least Squares for Straggler Mitigation in Distributed Recommendations}
\author{
\IEEEauthorblockN{Siyuan Wang\IEEEauthorrefmark{1}, Qifa Yan\IEEEauthorrefmark{2}, Jingjing Zhang\IEEEauthorrefmark{3}, Jianping Wang\IEEEauthorrefmark{1}, Linqi Song\IEEEauthorrefmark{1}}
\IEEEauthorblockA{\IEEEauthorrefmark{1}City University of Hong Kong,  \IEEEauthorrefmark{2}University of Illinoise at  Chicago, \IEEEauthorrefmark{3}Fudan University\\
Email: sywang34-c@my.cityu.edu.hk,  qifay2014@163.com, jingjingzhang@fudan.edu.cn, \{jianwang, linqi.song\}@city.edu.hk
}
}
\date{November 2020}
\begin{document}
\maketitle

% \graphicspath{{./figures/}}

\begin{abstract}
Matrix factorization is an important representation learning algorithm, e.g., recommender systems, where a large matrix can be factorized into the product of two low dimensional matrices termed as latent representations. 
This paper investigates the problem of matrix factorization in distributed computing systems with stragglers, those compute nodes that are slow to return computation results. A computation procedure, called coded Alternative Least Square (ALS), is proposed for mitigating the effect of stragglers in such systems.  The coded ALS algorithm iteratively computes two low dimensional latent matrices by solving various linear equations, with the Entangled Polynomial Code (EPC) as a building block. We theoretically characterize the maximum number of stragglers that the algorithm can tolerate (or the recovery threshold) in relation to the redundancy of coding (or the code rate). In addition, we theoretically show the computation complexity for the coded ALS algorithm and conduct numerical experiments to validate our design.  

%Recommender systems are adopted by many companies and have a strong impact on everyone's life in modern society. Collaborative filtering, which involves large amount of matrix decomposition, is of great importance in recommender system.  Alternating Least Squares(ALS) is an effective method to tackle the matrix decomposition problem. With the growing of user's data, the power of distributed learning is needed that computation tasks should be divided into subtasks and be distributed on multiple compute node. In such distributed system, however, because of hardware configurations or network delays, some workers that cannot return the results on time, also known as stragglers, could affect the efficiency of distributed algorithms. To tackle this, in this work, we design an distributed implementation of ALS algorithm and a stragglers-mitigate coding scheme using coding schemes to eliminate the impacts of stragglers. This approach divide the data matrix into $h^2$ batches  and encodes the batches as $W$ new batches to be processed on $w$ workers respectively.

\end{abstract}

\section{Introduction}
Matrix factorization is one of the most successful algorithms for many machine learning tasks  \cite{koren2009matrix}. For example, recommender systems have played an increasingly important role in the field of Internet business in recent years. Companies such as Amazon and Alibaba have used recommender systems to promote sales. Netflix, HBO, and YouTube have also used video recommender systems to recommend videos to target users. 
Since the Netfilx Prize competition held by Netflix \cite{bennett2007netflix}, the accuracy of recommendations have been greatly improved by matrix factorization algorithms.

With large amounts of data available nowadays, distributed computation is an important approach to deal with large scale data computations.  Straggler nodes are one of the most prominent problem in distributed computing systems \cite{karakus2017straggler,data2018data, li2018polynomially,bitar2019stochastic,ozfatura2019speeding,maity2019robust}.
Straggler is a node that runs much slower than others, which may be caused by various software or hardware issues such as hardware reliability, or network congestion.
As a result, straggler mitigation in distributed matrix multiplication - a basic building block for many machine learning algorithms - is crucial and has been extensively studied in the literature. Among them, coding techniques have attracted more attention recently in the information theory community, for example, Entangled Polynomial Code (EPC) \cite{yu2020straggler}.

In this paper, we investigate the problem of large scale matrix factorization through distributed computation systems. Consider a data matrix $R$ of size $m\times n$, where the dimensions $m$ and $n$ are typically very large so that each individual computing node can only deal with computations over matrices with dimensions much smaller than $\min\{m,n\}$. We aim to factorize $R$ approximately into the product $UV^\top$, where the dimensions of $U$ and $V$ are $m\times d$ and $n\times d$ respectively for some latent dimension $d\ll \mathrm{rank}(R)\leq \min\{m,n\}$.  
The factorization of $R$ can be formalized as minimizing $||R-UV^\top||_2^2$ for some $U,V$, where $||\cdot||_2$ is the Frobenius norm. 

Alternating Least Squares (ALS) is an efficient iterative algorithm to find out a solution by updaing $U$ and $V$ alternatively, where in each iteration, the algorithm  updates $U$ with the current estimate of $V$ and then updates $V$ with the updated estimate of $U$. 
We propose a distributed implementation of the ALS algorithm, with the ability to tolerate straggler nodes. In distributed ALS, matrix multiplication is a key building block and we adopt the EPC as a means to realize matrix multiplication, however, making special tailoring to the ALS algorithms where multiple matrix multiplications are involved at each iteration. To speed up the iteration, we first obtain the formulas that update $U$ from the current estimate of $U$ or  update $V$ from the current estimate of $V$. 
Based on the new update formulas, we only need to update either $U$ or $V$, since the estimates of $U$ and $V$ are connected by the original ALS formulas. 

Therefore, we propose a coded ALS framework as follows:
\begin{enumerate}
    \item Pre-computation: computing the transformed data matrix $RR^\top$ (or $R^{\top}R$) through the distributed computing system;
    \item Iterative computation: update $U$ (or $V$) through the distributed computing system;
    \item Post-computation: compute the estimate of $V$ (or $U$) from the estimate of $U$ (or $V$).
\end{enumerate}

In this computation framework, the bottleneck happens in the iterative computation phase, as the pre- and post- computations only need to be carried out once. 
In the iterative computations, both $U$ and $V$ are partitioned into submatrices along the larger dimension (i.e., $m$ or $n$), and the transformed data matrix $RR^\top$ or $R^\top R$ is  partitioned in both row and column dimensions and stored at the workers in coded form. We show that, with given partition parameters, the recovery threshold to compute matrix multiplications using EPC code is optimal among all linear codes (for matrix multiplication).  We characterize the relationship between the coding redundancy and the recovery threshold. In addition, we provide the computation complexity analysis for the proposed coded ALS algorithm. Finally, we conduct numerical
experiments to validate our proposed design.

\subsection{Related Work}
The slow machine problem has been existed in distributed machine learning for a long time\cite{ho2013more}. To tackle this, many different approaches have been proposed.In synchronous machine learning problems, solutions using speculative executions \cite{dean2008mapreduce,zaharia2008improving}. However, this types of methods need much more communication time and thus perform poor.

Adding redundancy is another effective way to cope with straggler problems. With each worker bear more information than it was supposed to, the final result could be recovered by these newly added extra information. In \cite{lee2017speeding}, the idea of using coded to solve straggler problems in distributed learning tasks. However, this work only focus on matrix multiplication and data shuffling. Then more and more research have been put into this area. One typical type of approach is 
data encoding, where the encoded data is stored in different workers. Works like \cite{lee2017speeding,data2018data,karakus2017straggler} encodes data as the linear combination of original data and recover the result according to the encoding matrix.

Another type is to encode the intermediate parameter of this code. A typical type of this coding is the gradient coding\cite{tandon2017gradient}. Many gradient based methods have been proposed within these years like \cite{neighbors1984draco,halbawi2018improving,maity2019robust}. However, these works focus only on gradient based distributed learning tasks.

\cite{dutta2018unified} Using coding in iterative matrix multiplication, which shares a similar application scenario with this paper.

\subsection{Statement of Contributions}
In this work, we make the following contributions.
1) Propose a coding scheme for large scale matrix decomposition problem to help with the recommender system.
2) Analysis of the complexity of this scheme and the running time of the coded distributed computation.
3) Solve the problem that the data partitions are too big when the numbers of columns and rows of the data matrix are both large.

\paragraph*{Notation} Throughout the paper, we use $[x]$ to denote the set $\{1,2,\ldots,x\}$, where $x \in \mathbb{N}_{+}$ is a positive integer. We denote by $||X|| = (\sum_{i \in [m],j \in[n]}X_{ij}^2)^{1/2}$ the Frobenius norm of the matrix $X \in \mathbb{R}^{m\times n}$. 
For easy of presentation, we will simply write $||X||_F$ as $||X||$ when there is no confusion.
\section{Problem Formulation}
\subsection{Matrix Factorization via ALS}
Given a data matrix $R \in\mathbb{R}^{m\times n}$ and a latent dimension $d$, we consider the matrix factorization problem to learn the $d$ dimensional representations $U\in \mathbb{R}^{m \times d}$ and $V\in \mathbb{R}^{n \times d}$ as follows:
\begin{equation}
\minimize\limits_{U\in \mathbb{R}^{m \times d}, V\in \mathbb{R}^{n \times d}} ||R-UV^\top||^2.
\end{equation}

For example, the matrix factorization can be used for the recommendation problem, where $m$ represents the number of users; $n$  represents the number of items; $(i,j)$ entry of the data matrix $R_{i,j}$  represents the rating (or preference) of user $i\in[m]$ to item $j\in[n]$. This rating can be approximated by the inner product of the latent vector $u_i$ of user $i$ and the latent vector $v_j$ of item $j$, where $u_i$ and $v_j$ are the $i$-th row of $U$ and the $j$-th row of $V$, respectively. The problem is to find these representations $u_i$ and $v_j$, so as to minimize the differences $\sum_{u_i,v_j}||R_{ij} - u_iv_j^\top||^2$. Note that parameters $m,n$ are often large and the latent dimension $d\ll \mathrm{rank}(R)\leq\min\{m,n\}$. 

Since this matrix factorization problem is non-convex, it is in general not easy to find the optimal solution. A well known iterative algorithm to solve this problem is the ALS algorithm, which iterates between optimizing $\sd{U}$ for given $\sd{V}$ and optimizing $\sd{V}$ for given $\sd{U}$. The update formulas for $\sd{U}$ and $\sd{V}$ are given as follows:
\begin{IEEEeqnarray}{rCl}
U^{(t+1)} &=& RV^{(t)} \Big( {V^{(t)}}^\top {V^{(t)}} \Big) ^{-1}\label{fr:updateu}\\
V^{(t+1)} &=& R^\top U^{(t+1)} \Big({U^{(t+1)}}^\top U^{(t+1)} \Big)^{-1}\label{fr:updatev}
\end{IEEEeqnarray}
where $\sd{U}^{(t)}$ and $\sd{V}^{(t)}$ are the estimates of $\sd{U}$ and $\sd{V}$ in the $t$-th iteration. In (\ref{fr:updateu}) $V$ is fixed, $U$ is updated and fixed to be used in (\ref{fr:updatev}), updating $V$ alternatively. 

\subsection{Distributed Matrix Factorization with Stragglers}

For large dimensions $m,n$, the updates in \eqref{fr:updateu} and \eqref{fr:updatev} are unpractical in a single computation node. 
We consider solving the matrix factorization problem via a `master-worker' distributed computing system with a master and $W$ workers. There may be some straggling workers (i.e., stragglers) among these workers which may perform the calculations slow and affect the system performance. 

We consider a coding aided framework to solve this matrix factorization problem.

$\bullet$ {\bf Data matrix encoding and distribution}: we first encode the data matrix $R$ to another matrix $\tilde R$ via some encoding function and the encoded data matrix will be distributed among workers. For worker $w$, the encoding and data distribution can be represented as $\tilde R_w = \phi_w(R)$ via some function $\phi_w()$.

$\bullet$ {\bf Iterative calculation and model aggregation}: the calculation is carried out in an iterative manner.  At each round $\tau$, each worker aims to calculate the model parameters and send model parameters to the master, namely, $\theta_w^{\tau+1} = {f}^{\tau}_w(\tilde R_w,\theta^{\tau})$, where ${f}^{\tau}_w()$ is the local computation function of worker $w$ at round $\tau$. However, there are a set of stragglers $W_s^{\tau}\subseteq W$ among these workers that either cannot calculate the result in time or cannot make successful transmissions. The server aggregates these model parameters from non-straggling workers in $W\backslash W_s^{\tau}$ to get a new model, namely, $\theta^{\tau+1} =  g^\tau(\{\theta_w^{\tau+1}\}_{w \in W\backslash W_s^{\tau}},\theta^{\tau})$, where $g^\tau()$ is the decoding and aggregation function at round $\tau$. After that, the server returns the updated parameter to all workers via a lossless broadcast channel. 

{\bf Our aim is to design a coded ALS scheme $\{\phi_w, f^{\tau}_w,g^\tau\}$ to solve the distributed matrix factorization problem\footnote{Here, we mean the coded ALS scheme achieves the same computation result as the centralized counterpart in Eqs.~\eqref{fr:updateu} and \eqref{fr:updatev}.} when there are no more than $s$ stragglers among the $W$ workers ($||W_s(\tau)||\leq s,\forall \tau$).}

Moreover, we would like to study how the coding scheme performs and the computation complexity with respect to the straggler mitigation capability $s$. In particular, let $\mathrm{size}(A)$ denotes the the number of elements in matrix $A$. We define the redundancy of coding (coding rate) $\mu$ as the ratio between the coded matrix size and the original data matrix size $\mu = W*\mathrm{size}(\tilde R_w)/\mathrm{size}(R)$. 

We then ask {\bf what is the relationship between $s$ and $\mu$ and how this further affects the computation complexity.}

\section{Distributed Computation of ALS Algorithm}
In this section, we will present our framework to implement the ALS algorithm in distributed computing systems.

\subsection{Preliminary: Entangled Polynomial Code}
%subsubsection{Alternating Least Squares Algorithm}
%Data matrix contains  elements that represent  scores of a user towards an item. Let $\sd{R}$ be the data matrix, and $d$ be a positive integer such that $d\ll\mathrm{rank}(\sd{R})\leq\min\{m,n\}$, which is used as the dimension of latent vector. The objective is to find $\sd{U}\in\sd{R}^{m\times d},\sd{V}\in\sd{R}^{n\times d}$ such that $\sd{U},\sd{V}$ minimize
%\begin{IEEEeqnarray}{c}
%||\sd{R}-\sd{U}\sd{V}^\top||_2^2.
%\end{IEEEeqnarray}
%A well known iterative algorithm to solve this problem is Alternative Least Square (ALS) algorithm, which iterates between optimizing $\sd{U}$ for given $\sd{V}$ and optimizing $\sd{V}$ for given $\sd{U}$. The update formulas for $\sd{U}$ and $\sd{V}$ are given by the least square estimation formula, i.e., 
%\begin{IEEEeqnarray}{rCl}
%U^{(t+1)} &=& RV^{(t)} \Big( {V^{(t)}}^\top {V^{(t)}} \Big) ^{-1}\\
%V^{(t+1)} &=& R^\top U^{(t+1)} \Big({U^{(t+1)}}^\top U^{(t+1)} \Big)^{-1}
%\end{IEEEeqnarray}
%where $\sd{U}^{(t)}$ and $\sd{V}^{(t)}$ are the estimates of $\sd{U}$ and $\sd{V}$ in the $t$-th iteration. For large dimensions $m,n$, the updates in \eqref{fr:updateu} and \eqref{fr:updatev} are unpractical in a single computation node. 

Entangled Polynomial Code (EPC) \cite{yu2020straggler} is an efficient linear code \cite[Definition 1]{yu2020straggler} to compute the large scale matrix multiplication  in distributed computation systems with stragglers.
%It is a type of \emph{linear code} for matrix multiplication:

\iffalse
\begin{definition}[Linear Code, \cite{yu2020straggler}] For a distributed matrix multiplication problem of computing $A^{\top}B$ using $W$ workers, we say a computation is a linear code with parameters $(p,q,r)$, where $\sd{A}$ and $\sd{B}$ are divided into the multiple  submatrices
such that the encoding functions of each worker $w$ can be written as
\begin{IEEEeqnarray}{c}
\widetilde{A}_w=\sum_{j,i} A_{j,i} a_{j,i,w},\quad \widetilde{B}_w=\sum_{j,k} B_{j,k}b_{j,k,w},
\end{IEEEeqnarray}
for some coefficients $\{a_{j,i,w}\}$ and $\{b_{j,k,w}\}$, and the decoding function can be written as 
\begin{IEEEeqnarray}{c}
\widetilde{C}_{i,k}=\sum_{w\in\mathcal{K}} \widetilde{C}_w c_{i,k,w}
\end{IEEEeqnarray}
for some coefficients $\{c_{i,k,w}\}$ for any active set $\mathcal{K}$, where $ \widetilde{C}_w$ is the response of worker $w$, and $\widetilde{C}_{i,k}$ is the estimate of the block $\sum_{j}A_{j,i}^{\top}B_{j,k}$.
\end{definition}
\fi

The entangled polynomial code computes $\sd{A}^\top \sd{B}$ via $W$ distributed worker nodes. Each worker node stores a coded sub-matrix based on the polynomials
\begin{IEEEeqnarray}{rCl}
\widetilde{\sd{A}}(x) &=&\sum_{j=0}^{r-1} \sum_{i=0}^{p-1} \sd{A}_{j, i} x^{j+i r}, \label{eqn:Ax}\\
\widetilde{\sd{B}}(x) &=&\sum_{j=0}^{r-1} \sum_{k=0}^{q-1} \sd{B}_{j, k} x^{r-1-j+k r p},
\label{eqn:Bx}
\end{IEEEeqnarray}
Specially, let $x_1,\ldots,x_W$ be $W$ distinct numbers in $\mathbb{R}$. Each worker $w\in[W]$ stores $\widetilde{\sd{A}}(x_w)$ and $\widetilde{B}(x_w)$, and computes and returns $\widetilde{\sd{A}}(x_w)^\top\widetilde{\sd{B}}(x_w)$. It was shown in \cite{yu2020straggler} that all the sub-matrices of the product $\sd{A}^\top\sd{B}$, i.e., 
\begin{IEEEeqnarray}{c}
C_{i,k}=\sum_{j=1}^rA_{j,i}^\top B_{j,k},\quad i\in[p],k\in[q]
\end{IEEEeqnarray}
are enclosed in the polynimal
$
\widetilde{\sd{C}}(x)=\widetilde{\sd{A}}(x)^\top\widetilde{\sd{B}}(x)
$,
which has degree $pqr+r-2$. As a result, the master can decode via interpolation 
from any responses of $K$ worker nodes, where
\begin{IEEEeqnarray}{rcl}
K=rpq+r-1
\end{IEEEeqnarray}
Given the parameters $p,q,r$, the $K$ of 
responses that the master needs to decode is called \emph{recovery theoreshold}.
It was showed in \cite[Theorem 2]{yu2020straggler}  that, when $p=1$ or $q=1$, the EPC achieves best recovery threshold among all \emph{linear code}.

%With Entangled Polynomial Code, only $pMN+p-1$ non-stragglers could suffice to recover the product. 

%To implement the ALS in such a distributed computing system, the matrix $\mathbf{R}$ is partitioned  into  $P\times Q$ equal-size sub-matrices for some positive integers\footnote{We implicitly assume that $p|m,q|n$} $P$ and $Q$, i.e., 

\subsection{Direct Update Formulas}
Unlike the traditional way (\ref{fr:updateu}) and (\ref{fr:updatev}) to update $U$ and $V$, here we choose to update either $U$ and $V$ only based on the following new update formulas. Note that here each iteration consists of several rounds.

\begin{lemma}\label{lemma:a}
In ALS algorithm, let $\sd{U}^{(t)},\sd{V}^{(t)}$ be the estimates of $\sd{U}$ and $\sd{V}$ in the $t$-th iteration, for each $t=1,2,\ldots T$,  
\begin{IEEEeqnarray}{rCl}
\sd{V}^{(t+1)}=\sd{R}^\top\sd{R}\sd{V}^{(t)}(\sd{ V}^{(t)\top}\sd{R}^\top\sd{R} \sd{V}^{(t)})^{-1} \sd{V}^{(t)\top} \sd{V}^{(t)}\label{fr:pr1recuisiveV}\\
\sd{U}^{(t+1)}=\sd{R}\sd{R}^\top\sd{U}^{(t)}(\sd{ U}^{(t)^\top}\sd{R}\sd{R}^\top \sd{U}^{(t)})^{-1} \sd{U}^{(t)\top}\sd{U}^{(t)}\label{fr:pr1recuisiveU}
\end{IEEEeqnarray}
\end{lemma}
\begin{IEEEproof}
E.q. \eqref{fr:pr1recuisiveV} can be obtained by plugging E.q. \eqref{fr:updateu} into E.q. \eqref{fr:updatev}, and E.q. \eqref{fr:pr1recuisiveU} can be  obtained by plugging E.q. \eqref{fr:updatev} (with the subscript $(t+1)$ replaced by $(t)$) into  E.q. \eqref{fr:updateu}.
For detailed proof, please refer appendix \ref{ap:singleside}

\end{IEEEproof}

\iffalse
Our proposed computation framework is based on the following observations:
\begin{enumerate}
    \item The equations \eqref{fr:pr1recuisiveV} and \eqref{fr:pr1recuisiveU} compute $\sd{V}^{(t+1)}$ from $\sd{V}^{(t)}$ and  $\sd{U}^{(t+1)}$ from $\sd{U}^{(t)}$ respectively;
\item The data matrix $R$ does appear in the update formulas other than in $R^\top R$ in \eqref{fr:pr1recuisiveV}, and $RR^\top$ in \eqref{fr:pr1recuisiveU}, and both $R^\top R$ and $RR^\top$ are independent of the iteration number $t$; 
\item The estimates of $U$ and $V$ are linked by \eqref{fr:updateu} and \eqref{fr:updatev}.  
\end{enumerate}
Based on the above observations, we propose to compute as follows: \fi
The main computation iterations either
update the estimate of $\sd{V}$ according to \eqref{fr:pr1recuisiveV} or
update the estimate of $\sd{U}$ according to \eqref{fr:pr1recuisiveU} instead of both in traditional ALS. Before the iteration, the matrix $R^{\top}R$ or $RR^\top$ is computed as a pre-computation to update $\sd{V}$ or $\sd{U}$. After obtaining the estimate of $V$ or $U$, a post-computation is used to compute the other factor via the relations \eqref{fr:updateu} or \eqref{fr:updatev}.

\subsection{Distribute ALS Algorithm}
In this subsection, we describe our algorithm in detail. The whole computation consists of three phases. 
\subsubsection{Pre-computation} If $m\geq n$, the master aims to compute 
$R^{\top}R
$
in order to update the estimate of $V$ according to \eqref{fr:pr1recuisiveV}; or else, the master aims to compute
$
RR^{\top}
$
according to the estimate of $U$ according to \eqref{fr:pr1recuisiveU}.
This can be done by using the standard EPC code, with appropriate parameters $p,q,r$. For clarity, in the following, we will denote matrix to be computed by $D$, and the factor that to be updated by $B$  i.e., 
\begin{IEEEeqnarray}{rCl}
D&=&\left\{\begin{array}{ll}
     R^\top R,&\mathrm{if}~m\geq n  \\
    RR^\top, & \mathrm{if}~m<n
\end{array}\right.,\label{eq:def:D}\\
B&=&\left\{\begin{array}{ll}
V,&\mathrm{if}~m\geq n\\
U,&\mathrm{if}~m<n
\end{array}
\right..
\end{IEEEeqnarray}
To have to smaller size of data to compute, we choose $D$ to be smaller one within $R^TR$ and $RR^T$. We choose $B$ in a similar way.
Notice that, by \eqref{fr:pr1recuisiveV} and \eqref{fr:pr1recuisiveU}, let $B^{(t)}$ be the estimate in the $t$-th iteration, then the update formula is 
\begin{IEEEeqnarray}{c}
B^{(t+1)}=DB^{(t)}\big(B^{(t)\top}DB^{(t)}\big)^{-1}B^{(t)\top}B^{(t)},\label{B:update}
\end{IEEEeqnarray}
where $D$ is an $l\times l$ symmetric matrix given by \eqref{eq:def:D}, $l=\min\{m,n\}$, and $B^{(t)}$ is of size $l\times d$.  
\subsubsection{Iterative Computation of $B$}

To implement the ALS algorithm in the distributed computing system, the matrix $D$ is partitioned into $h^2$ equal-size submatrices, each of size $\frac{l}{h}\times \frac{l}{h}$ for some positive integer $h$, i.e., 
\begin{IEEEeqnarray}{c}
\sd{D}=\left[\begin{array}{cccc}
\sd{D}_{0,0} & D_{0,1}&\ldots&\sd{D}_{0,h-1}\\
\sd{D}_{1,0}&D_{1,1}&\ldots&\sd{D}_{1,h-1}\\
\vdots&\vdots &\ddots&\vdots\\
\sd{D}_{h-1,0} &\sd{D}_{h-1,1}&\ldots&\sd{D}_{h-1,h-1}\\
\end{array}
\right]\label{partition:D}
\end{IEEEeqnarray}
In accordance with the partition in \eqref{partition:D}, $B^{(t)}$ is partitioned into $h$ equal-size sub-matrices of size $\frac{l}{h}\times d$, i.e., 
\begin{IEEEeqnarray}{c}
B^{(t)}=\left[\begin{array}{c}
B_0^{(t)}\\
\vdots\\
B_{h-1}^{(t)}
\end{array}\right],\quad \forall\, t\geq 0. \label{eqn:Bt:partition}
\end{IEEEeqnarray}

The bottleneck of updating the estimate of $B$ according to \eqref{B:update} is three matrix product computations: $DB^{(t)}$, $B^{(t)\top}DB^{(t)}$ and $B^{(t)\top} B^{(t)}$, and the product operation between the matrics $DB^{(t)}$ and $\big(B^{(t)\top}DB^{(t)}\big)^{-1}B^{(t)\top} B^{(t)}$,
which need to be computed at the distributed worker nodes\footnote{Notice that, as the dimensions of the matrix  $B^{(t)\top}DB^{(t)}$ and $B^{(t)\top}B^{(t)}$ are both  $d\times d$, the inverse operation in $\big(B^{(t)\top}DB^{(t)}\big)^{-1}$  and the product operation between $\big(B^{(t)\top}DB^{(t)}\big)^{-1}$ and $B^{(t)\top}B^{(t)}$ can be calculated at the master.}.

For clarity, we define the following polynomials:
\begin{IEEEeqnarray}{c}
\widetilde{D}(x)=\sum_{j=0}^{h-1}\sum_{i=0}^{h-1}D_{j,i}x^{j+ih}
\end{IEEEeqnarray}
where $\tilde D$ represents the encoded version of matrix $D$.

For any $C\in\mathbb{R}^{l\times d}$, partition $C$ in the same manner as in \eqref{eqn:Bt:partition}, i.e., $C=[C_0^\top,\ldots,C_{h-1}^{\top}]^\top$ where $C_i\in\mathbb{R}^{\frac{l}{h}\times d}$, define
\begin{IEEEeqnarray}{rCl}
f_{\mathrm{L}}(C,x)&=&C_0+C_1x+\ldots+C_{h-1}x^{h-1},\\
f_{\mathrm{R}}(C,x)&=&C_h+C_{h-1}x+\ldots+C_0x^{h-1}.
\end{IEEEeqnarray}

% 另外，这个C在之前的(9)式中就已经用到了。

Let $x_1,x_2,\ldots,x_W$ be $W$ distinct real numbers. The system operates as follows:
 
Initially, $B^{(0)}$ is generated randomly according to some continuous distribution.  The master sends $\widetilde{D}(x_w)$, $f_{\mathrm{L}}(B^{(0)},x_w)$ and $f_{\mathrm{R}}(B^{(0)},x_w)$ to the worker $w$ for each $w\in[W]$.
 
In each iteration $t=0,1,\ldots,T$
\begin{enumerate}
\item Each  worker $w\in[W]$  computes  
$
\widetilde{D}(x_w)f_{\mathrm{R}}(B^{(t)},x_w)$ and $f_{\mathrm{L}}(B^{(t)},x_{w})^{\top}f_{\mathrm{R}}(B^{(t)},x_w)$,
then
 responds the results to the master;
\item By EPC, receiving any $h^2+h-1$ results among 
\begin{IEEEeqnarray}{c}
\big\{\widetilde{D}(x_w)f_{\mathrm{R}}(B^{(t)},x_w):w\in[W]\big\},\label{eq:response:E}
\end{IEEEeqnarray}
the master can decode the matrix 
\begin{IEEEeqnarray}{c}
E^{(t)}\triangleq DB^{(t)}.\label{eq:Et}
\end{IEEEeqnarray}
Receiving any $2h-1$ results among 
\begin{IEEEeqnarray}{c}
\{f_{\mathrm{L}}(B^{(t)},x_{w})^{\top}f_{\mathrm{R}}(B^{(t)},x_w):w\in[W]\},\label{eq:response:F}
\end{IEEEeqnarray}
the master can decode the matrix
\begin{IEEEeqnarray}{c}
F^{(t)}\triangleq B^{(t)\top}B^{(t)}.\label{eq:Ft}
\end{IEEEeqnarray}
The master then sends $f_{\mathrm{R}}(E^{(t)},x_w)$ to worker $w$ for each $w\in[W]$;
\item Each worker $w\in[W]$ replace the matrix $f_{\mathrm{R}}(B^{(t)},x_w)$ with the matrix $f_{\mathrm{R}}(E^{(t)},x)$. Then it computes $f_{\mathrm{L}}(B^{(t)},x_w)^{\top}f_{\mathrm{R}}(E^{(t)},x_w)$ and sends the result to the master.  
\item Receiving any $2h-1$ responses among
\begin{IEEEeqnarray}{c}
\big\{f_{\mathrm{L}}(B^{(t)},x_w)^{\top}f_{\mathrm{R}}(E^{(t)},x_w):w\in[W]\big\},\label{eqn:response:BE}
\end{IEEEeqnarray}
the master decodes $B^{(t)\top}E^{(t)}$, and then the master computes
\begin{IEEEeqnarray}{c}
G^{(t)}=\big(B^{(t)\top}E^{(t)}\big)^{-1}F^{(t)},\label{eq:Gt}
\end{IEEEeqnarray}
and sends $G^{(t)}$ to all the worker nodes.
\item Each worker $w\in[W]$ computes 
\begin{IEEEeqnarray}{c}
f_{\mathrm{R}}(E^{(t)},x_w) G^{(t)}=f_{\mathrm{R}}(E^{(t)}G^{(t)},x_w)
\end{IEEEeqnarray}
and response $f_{\mathrm{R}}(E^{(t)}G^{(t)},x_w)$ to the master.
\item With any $h$ responses among
\begin{IEEEeqnarray}{c}
\{f_{\mathrm{R}}(E^{(t)}G^{(t)},x_w):w\in[W]\},\label{eqn:response:BG}
\end{IEEEeqnarray}
the master decodes
\begin{IEEEeqnarray}{c}
B^{(t+1)}=E^{(t)}G^{(t)}.\label{eq:BtPlus}
\end{IEEEeqnarray}
Then the master sends $f_{\mathrm{L}}(B^{(t+1)},x_w)$ and $f_{\mathrm{R}}(B^{(t+1)},x_w)$ to  worker $w$ for each $w\in[W]$.
\item Each worker $w\in[W]$ updates $f_{\mathrm{L}}(B^{(t)},x_w)$ and $f_{\mathrm{R}}(B^{(t)},x_w)$ with $f_{\mathrm{L}}(B^{(t+1)},x_w)$ and $f_{\mathrm{R}}(B^{(t+1)},x_w)$ respectively, and then starts the $(t+1)$-th iteration. 
\end{enumerate}

The procedure iterates until $B^{(t)}$ converges. Now, we claim that the above iterative procedure is correct. 

\begin{figure}[H]
    \centering
    \includegraphics[width=\linewidth]{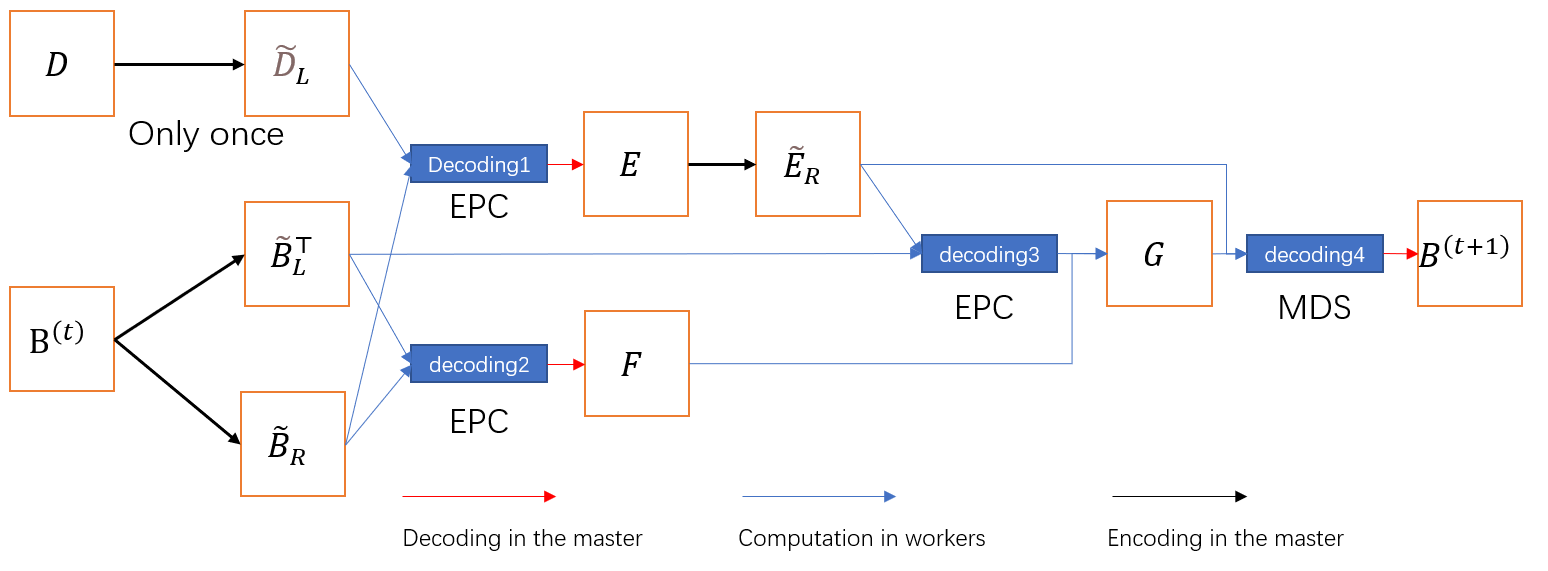}
    \caption{This figure describes the whole coding and decoding process of the algorithm. In each iteration, Firstly, $B$ is encoded into 2 copies that are $B_L$ and $B_R$, These two copies are used to create $E$ and $F$.  Finally $G$ will be created. With $G$ and $E$, we could obtain final $B^{(t+1)}$ in next iteration.}
    \label{fig:wholeprocess}
\end{figure}

In fact, it is easy to verify \eqref{B:update} by \eqref{eq:Et}, \eqref{eq:Ft},  \eqref{eq:Gt} and \eqref{eq:BtPlus}. We only need to show the following facts:
\begin{enumerate}
    \item [a)] With any $h^2+h-1$ responses in \eqref{eq:response:E}, the master can decode $E^{(t)}$. Notice that, the polynomials $\widetilde{D}(x)$ and $f_{R}(B^{(t)},x)$ are created according the $\eqref{eqn:Ax}$ and $\eqref{eqn:Bx}$ respectively, with parameters $p=h,r=h,q=1$. Thus, it directly follows from the result of EPC. 
    \item[b)] With any $2h-1$ responses in \eqref{eq:response:F} and \eqref{eqn:response:BE}, the master can decode $B^{(t)\top}B^{(t)}$ and $B^{(t)\top}E^{(t)}$ respectively. This directly follows by observing that the polynomials $f_{\mathrm{L}}(C,x)$ and $f_{\mathrm{R}}(C,x)$ are created according to  $\eqref{eqn:Ax}$ and $\eqref{eqn:Bx}$ respectively with parameters $p=q=1,r=h$. 
    \item[c)] With any $h$ responses in \eqref{eqn:response:BG}, the master can decode $E^{(t)}G^{(t)}$. This directly follows from the fact the polynomial $f_{\mathrm{R}}(C,x)$ has degree $h-1$, thus the polynomial $f_{\mathrm{R}}(E^{(t)}G^{(t)},x)$ can be obtained by Interpolation with any $h$ responses in  \eqref{eqn:response:BG}. 
\end{enumerate}
The whole process can be represented in Figure \ref{fig:wholeprocess}.
\begin{lemma}
\label{lm:threshold}
(Recovery threshold) The recovery threshold of the algorithm is given by %$K=h^2+h-1$.
%The whole process contains multiple matrix multiplication operation. 
%The recovery threshold for them is denoted by $K^{[1]}$, $K^{[2]}$ and $K^{[4]}$, corresponding to the 4 decoding process in Fig.\ref{fig:wholeprocess}. Then
\begin{IEEEeqnarray}{rcl}
K=h^2+h-1.
\label{fr:recoverythreshold}
\end{IEEEeqnarray}
\textit{Proof}: From the above descriptions a), b), c), the iteration involves four distributed matrix multiplication. The recovery threshold is given by the maximum recovery threshold of those updates, i.e., 
\begin{IEEEeqnarray}{rcl}
K=\max(h^2+h-1, 2h-1, h)=h^2+h-1.
\end{IEEEeqnarray}
%Thus we can get
%\begin{IEEEeqnarray}{rcl}
%K^{[1]}= h^2+h-1\\
%K^{[2]}=T^{[3]}= 2h-1
%\end{IEEEeqnarray}
%When $h>0$, it is obvious that $h^2+2h-1>2h-1>h$.

%Therefore, $K=K^{[1]}=h^2+h-1$.

\end{lemma}

\begin{remark}[Optimality of the Recovery Thresholds]
Notice that, with the given parameter $h$, from the results of EPC in \cite[Theorem 2]{yu2020straggler}, all the EPC code in a) and b) achieves the  optimal recovery threshold among all linear code for the corresponding matrix multiplication problems. For the calculation in c), it also achieves the optimal recovery threshold by simple cut-set bound.
\end{remark}
\subsubsection{Post-Computation} The master has obtained the final estimate of $B$, denoted by $\widetilde{B}$. The master first computes $\widetilde{H}=\widetilde{B}(\widetilde{B}^{\top}\widetilde{B})^{-1}$ as follows:
\begin{enumerate}
    \item The master sends $f_{\mathrm{L}}(\widetilde{B},x_w)$ and $f_{\mathrm{R}}(\widetilde{B},x_w)$ to worker $w$ for each $w\in[W]$;
    \item Each worker $w\in[W]$ computes $f_{\mathrm{L}}(\widetilde{B},x_w)^{\top}f_{\mathrm{R}}(\widetilde{B},x_w)$ and sends the results to the master;
\item With any $2h-1$ responses, the master decode 
\begin{IEEEeqnarray}{c}
\widetilde{F}=\widetilde{B}^{\top}\widetilde{B}
\end{IEEEeqnarray}
by EPC decoding. Then, it computes and sends $\widetilde{F}^{-1}$ to all the worker nodes;
\item Each worker computes
\begin{IEEEeqnarray}{c}
f_{\mathrm{L}}(\widetilde{B},x_w)\widetilde{F}^{-1}=f_{\mathrm{L}}(\widetilde{B}\widetilde{F}^{-1},x_w)
\end{IEEEeqnarray}
and sends it back to the master;
\item With any $h$ responses, the master decodes $\widetilde{H}=\widetilde{B}\widetilde{F}^{-1}$ by interpolating the polynomial $f_{\mathrm{L}}(\widetilde{B}\widetilde{F}^{-1},x)$.
\end{enumerate}
Then the master continue to obtain the estimates of $V$ and $U$ as follows:
\begin{enumerate}
    \item If $m\geq n$, the estimate of $V$ is given by $\widetilde{V}=\widetilde{B}$. The estimate of $U$ is given by 
    \begin{subequations}\label{post:update}
    \begin{IEEEeqnarray}{c}
    \widetilde{U}=R^{\top}\widetilde{H},
    \end{IEEEeqnarray}
    which computed with standard EPC code with appropriate parameters.
    \item If $m<n$, the estimate of $U$ is given by $\widetilde{U}=\widetilde{B}$, the estimate is given by 
    \begin{IEEEeqnarray}{c}
    \widetilde{V}=R\widetilde{H},
    \end{IEEEeqnarray}
    \end{subequations}
    which can be computed by standard EPC code with appropriate parameters. 
\end{enumerate}
\begin{remark}
Since the main computation load is in the iterative computation of $B$, we omit the details of computation of $D$  in \eqref{eq:def:D} and the computations in \eqref{post:update}. One convenient choice of the partition parameters is partition $R$ (if $m\geq n$) or $R^{\top}$ (if $m<n$) into $r\times h$ equal-size sub-matrices for some $r$, and partition $\widetilde{H}$ in the same form as in \eqref{eqn:Bt:partition}, so that the result of $D$ has the form  \eqref{partition:D}. Under such partitions, the EPC computation of \eqref{post:update} also achieves the optimal recovery threshold among all linear code since $q=1$ by \cite[Theorem 2]{yu2020straggler}.  
\end{remark}

\iffalse

Based on this, we could find that, if we compute either (\ref{fr:updateu}) or (\ref{fr:updatev}) twice in each iteration, the complexity could be close to  $O(m^2d)$ or $O(n^2d)$. If $m<n$ or $n<m$,  the \qf{fewer} one in $O(m^2d)$ or $O(n^2d)$ could be be less than $O(2mnd)$, which means a large amount of time could be saved by only updating $\mathbf U$ or $\mathbf V$.

This inference is not rigorous, but gives us an intuition of using the moment coding\cite{maity2019robust} to reduce the computation complexity. The computation of $\mathbf R^T\mathbf R$ or $\mathbf {RR}^T$ only need to be done once in the process.
\fi

%\input{4-Analysis}
\section{Main Results}

The following theorem characterizes the maximum number of stragglers that the algorithm can tolerate in relation to the coding redundancy. 
\begin{theorem}
 The relation between the recovery threshold $K$ (or the maximum number of stragglers that the coded ALS algorithm can tolerate $s = W -K$) and coding redundancy $\mu$ is given by
\begin{IEEEeqnarray}{rcl}
K=\frac{W}{\mu}+ \sqrt{\frac{W}{\mu}}-1, \quad \mu>1
\label{fr:storageRedundancy}
\end{IEEEeqnarray}
where $W\geq h^2$.
\end{theorem}
\begin{IEEEproof} 
Suppose each worker have a same size of data partitions of $\tilde R$ with $k$ elements.
According to \eqref{partition:D} and the fact that each worker holds one partition of $R$, and there are $h^2$ partitions in $R$, and the number of elements in $R$ is $h^2k$.
So that $\mu$ could be represented as
\begin{IEEEeqnarray}{rcl}
\mu=\frac{Wk}{h^2k}=\frac{W}{h^2}
\label{fr:storagerate}
\end{IEEEeqnarray}

From Lemma \ref{lm:threshold}, we know the relation between $K$ and $h$. By replacing $h$ in \eqref{fr:recoverythreshold} by \eqref{fr:storagerate}, we can get \eqref{fr:storageRedundancy}.
\end{IEEEproof}

\begin{theorem}
(Computation Complexity Analysis)
Given a distributed matrix factorization problem with $W$ workers, $m\times n$ matrix $R$, latent dimension $d$, and coding redundancy $\mu$, the computation complexity of the proposed coded ALS algorithm can be calculated as follows.

\begin{enumerate}
    \item\label{th:complexity:pre-computation} The pre-computation complexity (at master) is $O(\min\{mn^2,nm^2\})$ (one time).
    
    \item\label{th:complexity:computation} The computation complexity at each worker is  $O(\frac{n^2d\mu}{W})$ at each iteration\footnote{Each iteration involves both updates of $V$ and $U$}.
    
    \item\label{th:complexity:coding} The encoding and decoding complexities are $O(nd\sqrt{W\mu})$ and $O(nd(\frac{W}{\mu}+\sqrt{\frac{W}{\mu}}))$ at the master at each iteration. 
    
    \item\label{th:complexity:decoding} The decoding complexity, donotes the decoding procedure happened in the master, is $O(nd(h^2+h))$
\end{enumerate}

For the whole process proposed in this algorithm depicted in Fig.\ref{fig:wholeprocess}, where we use $T$ to represent the number of iterations, and $n$ to represent the number of columns and rows of $B$\footnote{Here we assume that $n<m$ for R}. The complexity for each stage of the algorithm can be represented as the following form:
\iffalse
\begin{enumerate}
    \item\label{th:complexity:computation} The computation complexity can be represented as  $O(\frac{Tn^2d}{h^2})$ in each worker.
    
    \item\label{th:complexity:encoding} The encoding complexity, which denotes the complexity to do the encoding procedure denoted as $f_L$ and $f_R$ taken place in the master, is $O(\frac{TndW}{h})$. 
    
    \item\label{th:complexity:decoding} The decoding complexity, donotes the decoding procedure happened in the master, is $O(nd(h^2+h))$.
\end{enumerate}
\fi

\end{theorem}

\begin{IEEEproof}
Consider the size of matrix multiplication, we can measure the computation complexity of different procedure of this proposed algorithm.
%See appendix in \cite{appendixISIT} for a full proof.
In the pre-computation stage, we only have a computation task of $R^TR$ or $RR^T$, and thus the the complexity is $O(\min{mn^2, nm^2})$.
In each worker, we only consider the first stage that is $E=DB$. The two matrices involved in the computation have size that are $\frac{n^2}{h^2}$ and $\frac{nd}{h^2}$. Replace $h$ by $h=\sqrt{W}{\mu}$, we will get the complexity of each worker.
For the encoding and decoding complexity, it is an weighted sum of matrices from each workers and data partitions. We consider the number of workers $W$ and number of data partitions $h$.
Finally, we could get the encoding and decoding complexity respectively.
\end{IEEEproof}
\iffalse
The computation of $E$ is the bottleneck of the whole process. Consider the size of both $D$ and $B$,
the computation complexity in each worker could be represented as $O(\frac{Tn^2d}{h^2})$ for the whole algorithm in each worker.

For encoding and decoding process taken place in the master, the bottle neck is the weighed sum of results from each workers. So that we consider the largest matrix involed in the encoding and decoding process. So that the encoding and decoding process could be represented as $O(\frac{TndW}{h})$ and $O(nd(h^2+h))$ respectively.
\fi

\begin{remark}
\upshape{(Number of partitions)}
In the proposed algorithm, to get a better partition method, in common configurations, e.g., when $W<100$. When $h^*=\lfloor \sqrt{W+\frac{3}{4}-s} \rfloor$, we will
achieve a relatively less expected computation time $\mathbb E[T_{cp}]$ when workers doing their computation task.

\end{remark}

By the definition of stragglers $s+K\leq W$ 
and the fact that $K=h^2+h-1$, we could find the largest $h$ that could be tolerated by the algorithm, which will therefore save a large amount of time in practical.
%See full proof please see \cite{appendixISIT}.

According to the definition of stragglers
\begin{IEEEeqnarray}{rcl}
s+K\leq W
\label{fr:upperbound}
\end{IEEEeqnarray}

There are 4 decoding procedure in the algorithm, shown in figure \ref{fig:wholeprocess}, the recovery threshold for the first step to get  $E$ is the largest which is 
\begin{IEEEeqnarray}{rcl}
K=f_K(h)=h^2+h-1
\label{fr:threshold}
\end{IEEEeqnarray}

In \textbf{Remark} \ref{pr:time_decrease}, we have discussed that the computation time $T_{cp}$ decreases as $h$ grows.
Therefore, to get smaller $T_{cp}$, we need to make $h$ greater.

It's easly known that $f_K(h)$ monotonously increased when $h\geq 1$, which is upper bounded by \eqref{fr:upperbound}.
Therefore, \eqref{fr:upperbound} and \eqref{fr:threshold} can be written as
\begin{IEEEeqnarray}{rcl}
h^2+h+s-(w+1)\leq 0
\end{IEEEeqnarray}
and therefore the best $h^*$ is got to make to make the computation time less.

\begin{remark}
\label{pr:time_decrease}
In worker's computation stage, when doing the multiplication of matrices,
the computation time $T_{cp}$ decreases as $h$ grows.
\end{remark}
We consider two impacts of this algorithm. The first impact is more partitions makes each worker share less proportion of data which speeds up the computation. The second impact is more partitions require more usable worker, which will make us get an bigger order statistics.  
See Appendix \ref{ap:decrease} for full explanation.

$\mu$ and $\sigma$ are determined by the performance of the computer, which is hard to measure in this setup.Intuitively, with larger $h$, each worker has smaller data to process, the total computation time could be less.

\section{Simulations}
In this section, we design a simulation experiment to measure the running time of our algorithm.

We conduct our simulations on synthetic data by adding noise into the product of two randomly initiated matrices $U$ and $V$. In the simulation, we set the parameters $m=2400$, $n=1500$, $d=200$ and run the simulations when the number of non-stragglers $k= W-s= 10,20,30,40,50$. We tested the computation time of the proposed algorithm of different number of data partitions and record the result in Table \ref{tb:simulation_result}.

\begin{table}[H]
\centering
\label{tb:simulation_result}
\caption{Computation time for different partitions} 
\begin{tabular}{|c|c|c|c|c|c|}\hline
\quad   & $h=2$      & $h=3$      & $h=4$      & $h=5$      & $h=6$      \\ \hline
$k=10$ & 7.59468  &      -    &      -    &     -    & -         \\ \hline
$k=20$ & 7.376128 & 3.952123 & 2.856271 &     -     & -         \\ \hline
$k=30$ & 6.581517 & 3.7818   & 2.729829 & 0.113088 &  -        \\ \hline
$k=40$ & 6.860791 & 3.744443 & 1.285898 & 0.008618 &  -        \\ \hline
$k=50$ & 6.852958 & 3.778484 & 0.937565 & 0.003492 & 0.000598 \\ \hline
\end{tabular}
\end{table}

In Table \ref{tb:simulation_result}, we could easily see the relationship between the computation time in workers and the number of partitions $h$ in data matrix $R$ for different number of non-straggling workers $k$. The minus sign in the table means there is no data, because the condition $h^2+h-1+s\leq W$ is not met. Intuitively, smaller partitions result in a faster computation time in workers although smaller partitions will lead to more workers involved in the computation and therefore increase the computation time. For the same $h$, larger $k$ could make the computation faster on the whole.
That's because more workers in total bring more  workers that are faster. Given the same number of workers, we can also see that when the coding redundancy $\mu$ increases ($h$ decreases), we need less computation time, indicating a tradeoff of straggler mitigation ability and computation complexity in the simulation results.

\iffalse
In Fig.\ref{fig:encoding_time}, we show the encoding time of different number of stragglers on cases when $p=2$ and $p=3$.

\begin{figure}[H]
    \centering
    \includegraphics[width=\linewidth]{figures/encoding_time.png}
    \caption{This figure compares encoding time when $p=2$ and $p=3$. of different number of stragglers. The encoding time seems to be a fixed number. We could see that when $p=3$, it needs more time to encode the original data.}
    \label{fig:encoding_time}
\end{figure}

\begin{figure}[H]
    \centering
    \includegraphics[width=\linewidth]{figures/decoding_time.png}
    \caption{This figure compare the decoding time of the cases when $p=2$ and $p=3$. The decoding time grows slightly with the number of stragglers. That's because the size of encoding matrix grows with $p$. According to (\ref{fr:decoding_time}) that is in proportion to $p^2+2p-1$, the decoding time when $p=3$ is almost twice the time when $p=2$.}
    \label{fig:my_label}
\end{figure}
\fi

\section{Conclusion}
We presented a distributed  implementation of the ALS algorithm, which solves the matrix factorization problem in a distributed computation system.  The procedure takes the advantage of the entangled polynomial code as a building block, which can resist stragglers. The relationship between the recovery threshold and the storage load is characterized. The simulation result indicates that with more workers and partitions of data matrix, we could obtain a shorter computation time, and thereby fully implemented the role of distributed learning.
%This paper gives a distributed implementation of the ALS algorithm and a coding scheme to mitigate the number of stragglers. Afterwards, we did experiments to show the effectiveness of the coding scheme. 
%However, there are still some problems that needs to be solved in the future. One of them is that in this problem we assume the data matrix $R$ is complete. However, in the real world, $R$ is often incomplete.  

\begin{appendices}
\section{}
\label{ap:decrease}
Let $T_{cp}$ represents the total time consuming in the encoding process, and thus
\begin{IEEEeqnarray}{rcl}
T_{cp} = T^{[1]}_{cp} + T^{[2]}_{cp} + T^{[3]}_{cp} + T^{[4]}_{cp}
\label{fr:total_time}
\end{IEEEeqnarray}
where $T^{[A]}_{cp}$ represents the $A$-th computation stage.
Because $F$, $B^T E$ and $B$ have the same size of matrix to do the multiplication, so the 
\begin{IEEEeqnarray}{rcl}
\mathbb E T^{[2]}_{cp} =\mathbb E T^{[3]}_{cp}
\end{IEEEeqnarray}
So that the expected encoding time can be represented as
\begin{IEEEeqnarray}{rcl}
\mathbb E T_{cp}= \mathbb E T_{cp}^{[1]} + 2 \mathbb E T_{cp}^{[2]} + \mathbb E T_{cp}^{[4]}
\end{IEEEeqnarray}
where $T_{cp}^{[A]} = \sum\limits_j^T T_{cp(t)}^{[A]}$, in which $t$ represents the $t$-th iteration of the algorithm and $T_{cp(t)}^{[A]}$ represents the total time consumed of stage $A$ at $t$-th iteration.

For each worker, The primary task is to compute  $E$, whose computation time is denoted as $T_{cp}^{[1]}$. The computation time for it is $\sum\limits_l^\frac{n^2d}{h^2}T_{u(l)}$ for each worker at each iteration. 
To compute $F$, $B^T E$ and $B$,denoted as $T_{cp}^{[2]}$,$T_{cp}^{[3]}$, $T_{cp}^{[4]}$, where each worker needs $\sum\limits_l^\frac{d^2n}{h}T_{u(l)}$ at each iteration, where  $T_{u(l)}\sim N(\mu_u, \sigma_u^2)$ represents the time to do a element-wise multiplication in one worker.

The computation time at each iteration can be represented as the following form.
\begin{IEEEeqnarray}{rcl}
T_{cp(t)}^{[A]} = \max\limits_{(i)}^{W-s} T_{cp(t)}^{[A]}, i\in[W]
\end{IEEEeqnarray}
where $\max\limits_{(i)}^{W-s} T_{cp(t)}^{[A]}$ denotes the $i$-th shortest time taken by the workers to return the result. 

The multiplication of matrices can be seen as element-wise operation in each worker. 
According to the \textit{Central Limit Theorem}, the $^{i}T_{cp(j)}^{[1]}$ can be represented as 

\begin{IEEEeqnarray}{rcl}
^{i}T_{cp(t)}^{[1]} \sim N(\frac{n^2d}{h^2}\mu_u,  \frac{n^2d}{h^2}\sigma_u^2)
\label{fr:distribution1}
\end{IEEEeqnarray}
where $^{i}T_{cp(t)}^{[1]}$ represents the time taken by worker $w_i$ in $t$-th iteration in stage 1.

The expected computation time can be represented as
\begin{IEEEeqnarray}{rcl}
\mathbb E T_{cp}^{[1]} = T \mathbb E[\max\limits_{(i)}^{W-s} T_{cp(t)}^{[1]}], i\in[W]
\end{IEEEeqnarray}

Similarly, the $\mathbb E T_{cp}^{[2]}$ can be represented as same form, and 

\begin{IEEEeqnarray}{rcl}
T_{cp(t)}^{[2]} \sim N(\frac{nd^2}{h}\mu_u,  \frac{nd^2}{h}\sigma_u^2)
\label{fr:distribution2}
\end{IEEEeqnarray}

According to the \cite{royston1982algorithm}, the approximation of expected value of the $r$-th order statistic can be represented as
\begin{IEEEeqnarray}{rcl}
E(r:n) \approx \mu_u + \Phi^{-1}(\frac{r-\alpha}{n-2\alpha+1})\sigma_u
\end{IEEEeqnarray}
where $\alpha=0.375$ and $\Phi(x)$ is the inverse function of the cumulative distribution function of standart normal distribution.

Because $n^2d>>nd2$, combining \eqref{fr:distribution1} and \eqref{fr:distribution2}, we know that $\mathbb E T_{cp}^{[1]}>>\mathbb E T_{cp}^{[i]}$ when $i\in\{2,3,4\}$ 

so that the computation time in \eqref{fr:total_time} could be written as
\begin{IEEEeqnarray}{rcl}
\mathbb E T_{cp}\approx T\mathbb E[\max\limits_{(i)}^{W-s} T_{cp(t)}^{[A]}]\\
\approx T(\frac{n^2d}{h^2}\mu_u + \Phi^{-1}(\frac{h^2+h-1-\alpha}{W-s-2\alpha+1})\frac{\sqrt{d}n}{h}\sigma_u )
\label{fr:est_T}
\end{IEEEeqnarray}

Let $\theta(h)$ represents the right hand side term in \eqref{fr:est_T} and then $\theta'(h)$ can be written as
\begin{IEEEeqnarray}{rcl}
\theta'(h)\!=\! -2\mu_u n^2\!dh^{-3}\!
+\!\frac{(2h\!+\!1)\sqrt{d}n\sigma_u}{\phi(\frac{h^2+h-1-\alpha}{W-s-2\alpha+1})h}\!
-\!2\Phi^{-\!1}\!(\!\frac{h\!^2\!+h\!-1\!-\!\alpha}{\!W\!-\!s\!-\!2\!\alpha\!+\!1}\!)\sqrt{d}n\sigma_u h^{-2}
\end{IEEEeqnarray}
where $\phi(x)$ denotes the probability density function of standard normal distribution.

Dividing both sides by $2n\sqrt{d}\sigma$, we could get
\begin{IEEEeqnarray}{rcl}
\theta_2'(h) = - \frac{\mu_u n \sqrt{d} h^{-3}}{\sigma_u}\! +\! \frac{1+\frac{1}{2h}}{\phi(\frac{h^2+h-1-\alpha}{W-s-2\alpha+1})}
\!- \!\Phi^{-1}\!(\frac{h\!^2\!+h\!-1\!-\!\alpha}{\!W\!-\!s\!-\!2\!\alpha\!+\!1})h^{-2}
\label{fr:simplfied_diff}
\end{IEEEeqnarray}

\begin{equation}
\begin{array}{llll}
\theta_2'(h) = - \frac{\mu_u n \sqrt{d} h^{-3}}{\sigma_u}\! +\! \frac{1+\frac{1}{2h}}{\phi(\frac{h^2+h-1-\alpha}{W-s-2\alpha+1})} \!- \!\Phi^{-1}\!(\frac{h\!^2\!+h\!-1\!-\!\alpha}{\!W\!-\!s\!-\!2\!\alpha\!+\!1})h^{-2}
\label{fr:simplfied_diff}
\end{array}
\end{equation}

So that we only need to \eqref{fr:simplfied_diff} to determine whether it's positive or negative.

In \eqref{fr:simplfied_diff}, let's use $\phi(\cdot)$ to represent $\phi(\frac{h^2+h-1-\alpha}{W-s-2\alpha +1})$ and use $\Phi(\cdot)$ to represent $\Phi(\frac{h^2+h-1-\alpha}{W-s-2\alpha +1})$. $\phi(\cdot)$ is a decreaseing function after $h\geq 1$. When $h=1$, $\phi(\cdot)$ reaches its maximum $\phi(\frac{1-\alpha}{100-2\alpha+1})$, which is less than 0.398 when $W-s \leq 100$; When $h=h_M$, where $h_M$ denotes the max $h$ such that $s+h^2+h-1\leq W$, $\phi(\cdot)$ reaches its minimum $\min\phi(\cdot)=\phi(\frac{h_M^2+h_M-1-\alpha}{W-s-2\alpha +1})\geq \phi(\frac{W-s-1-\alpha}{W-s-2\alpha +1})$ which is greater than 0.246 when $W-s\leq 100$.

Therefore, 
\begin{IEEEeqnarray}{rcl}
0.246\leq \phi(\cdot) \leq 0.398 \quad W-s\leq 100
\end{IEEEeqnarray}
meanwhile, $1+\frac{1}{2h}$ meets
\begin{IEEEeqnarray}{rcl}
1.056\leq 1+\frac{1}{2h} \leq 1.5
\end{IEEEeqnarray}
So that, the second term in \eqref{fr:simplfied_diff} meets
\begin{IEEEeqnarray}{rcl}
 2.6537 \leq \frac{1+\frac{1}{2h}}{\phi(\cdot)} \leq 6.0975
\end{IEEEeqnarray}

Let's use $\Phi^{-1}(\cdot)$ to represent $\Phi^{-1}(\frac{h^2+h-1-\alpha}{W-s-2\alpha +1})h^{-2}$. Then $\Phi^{-1}(\cdot)$ is increasing when $h\geq 1$. So the $\Phi^{-1}(\cdot)$ reaches its maximum when $h=h_M$
\begin{IEEEeqnarray}{rcl}
\max \Phi^{-1}(\cdot)
\leq \Phi^{-1}(\frac{W-s-\alpha}{W-s-2\alpha+1}) \leq \Phi^{-1}(0.98379)=2.14\quad W\leq 100
\label{fr:floor_Phi}
\end{IEEEeqnarray}
when $h=1$
\begin{IEEEeqnarray}{rcl}
\min\Phi^{-1}(\cdot)\geq\Phi^{-1}(\frac{1-\alpha}{W-s-2\alpha+1})
\geq -2.5 \quad W\leq 100
\label{fr:ceil_Phi}
\end{IEEEeqnarray}
Therefore, $-2.14\leq -\Phi^{-1}(\cdot)h^{-2}\leq 2.5$. In other words, the second and third term in \eqref{fr:simplfied_diff} are all bounded. If we can guarantee
\begin{IEEEeqnarray}{rcl}
\frac{\mu_u n \sqrt{d}}{h^3\sigma_u}\geq 8.5975\quad W\leq 100
\label{fr:request}
\end{IEEEeqnarray}
such that $\theta_2'(h)<0$, which suggests $\theta(h)$ decrease monotonously when $h\geq 1$, resulting in a less time of computation with a greater $h$.

\section{}
\label{ap:singleside}
\begin{IEEEeqnarray}{c}
\begin{split}\label{fr:pr1a}
\sd{V}^{(t+1)}=
\sd{R}^\top\sd{R} \sd{V}^{(t)} (\sd V^{(t)\top}\sd{V})^{-1}( (\sd{R}\sd{V}^{(t)}(\sd{V}^{(t)\top}\sd{ V}^{(t)})^{-1})^\top
({RV}( V^T V)^{-1}) )^{-1}
\end{split}
\end{IEEEeqnarray}
in which $( R V( V^T V)^{-1})^T$ could be writen as 
\begin{IEEEeqnarray}{c}
{( V^T  V)^{-1}}^T R^T V^T\label{fr:pr1invertU}
\end{IEEEeqnarray}
Because ${ A^{-1}}^T={ A^T}^{-1}$, and that $ V^T V$ is a symmetric matrix, (\ref{fr:pr1invertU}) could be $( V^T V)^{-1} R^T V^T$
With (\ref{fr:pr1invertU}) and (\ref{fr:pr1a}), we will get that
\begin{IEEEeqnarray}{rCl}
\begin{split}
 V^{(t+1)}=
 R^T {RV}( V^T V)^{-1}
(( V^T V)^{-1}
 V^T
 R^T
{RV}
( V^T V)
)^{-1}
\end{split}
\label{fr:pr1b}
\end{IEEEeqnarray}
Now we know that $ U^T U$ is invertible. So that the items within the rightmost bracket is invertible. From the fact that $({ABC})^{-1}= C^{-1} B^{-1} A^{-1}$, the RHS of \ref{fr:pr1b} could be
\begin{IEEEeqnarray}{c}
(( V^T V)
( V^T
 R^T
{RV})^{-1}
(( V^T V)^{-1})
)
\end{IEEEeqnarray}
and \ref{fr:pr1b} becomes
\begin{IEEEeqnarray}{rCl}
\begin{split}
 V^{(t+1)}=
 R^T {RV}( V^T V)^{-1}
(( V^T V)
( V^T
 R^T
{RV})^{-1}
(( V^T V)^{-1})
)\end{split}
\label{fr:pr1c}
\end{IEEEeqnarray}
Simplify (\ref{fr:pr1c}) we will finally get (\ref{fr:pr1recuisiveV}). Similarly, we can do the same procedure and then get (\ref{fr:pr1recuisiveU})
\end{appendices}
\bibliographystyle{IEEEtran}
\bibliography{references}{}

\begin{thebibliography}{10}
\providecommand{\url}[1]{#1}
\csname url@samestyle\endcsname
\providecommand{\newblock}{\relax}
\providecommand{\bibinfo}[2]{#2}
\providecommand{\BIBentrySTDinterwordspacing}{\spaceskip=0pt\relax}
\providecommand{\BIBentryALTinterwordstretchfactor}{4}
\providecommand{\BIBentryALTinterwordspacing}{\spaceskip=\fontdimen2\font plus
\BIBentryALTinterwordstretchfactor\fontdimen3\font minus
  \fontdimen4\font\relax}
\providecommand{\BIBforeignlanguage}[2]{{%
\expandafter\ifx\csname l@#1\endcsname\relax
\typeout{** WARNING: IEEEtran.bst: No hyphenation pattern has been}%
\typeout{** loaded for the language `#1'. Using the pattern for}%
\typeout{** the default language instead.}%
\else
\language=\csname l@#1\endcsname
\fi
#2}}
\providecommand{\BIBdecl}{\relax}
\BIBdecl

\bibitem{koren2009matrix}
Y.~Koren, R.~Bell, and C.~Volinsky, ``Matrix factorization techniques for
  recommender systems,'' \emph{Computer}, no.~8, pp. 30--37, 2009.

\bibitem{bennett2007netflix}
J.~Bennett, S.~Lanning \emph{et~al.}, ``The netflix prize,'' in
  \emph{Proceedings of KDD cup and workshop}, vol. 2007.\hskip 1em plus 0.5em
  minus 0.4em\relax New York, NY, USA., 2007, p.~35.

\bibitem{karakus2017straggler}
C.~Karakus, Y.~Sun, S.~Diggavi, and W.~Yin, ``Straggler mitigation in
  distributed optimization through data encoding,'' in \emph{Advances in Neural
  Information Processing Systems}, 2017, pp. 5434--5442.

\bibitem{data2018data}
D.~Data, L.~Song, and S.~Diggavi, ``Data encoding for byzantine-resilient
  distributed gradient descent,'' in \emph{2018 56th Annual Allerton Conference
  on Communication, Control, and Computing (Allerton)}.\hskip 1em plus 0.5em
  minus 0.4em\relax IEEE, 2018, pp. 863--870.

\bibitem{li2018polynomially}
S.~Li, S.~M.~M. Kalan, Q.~Yu, M.~Soltanolkotabi, and A.~S. Avestimehr,
  ``Polynomially coded regression: Optimal straggler mitigation via data
  encoding,'' \emph{arXiv preprint arXiv:1805.09934}, 2018.

\bibitem{bitar2019stochastic}
R.~Bitar, M.~Wootters, and S.~E. Rouayheb, ``Stochastic gradient coding for
  straggler mitigation in distributed learning,'' \emph{arXiv preprint
  arXiv:1905.05383}, 2019.

\bibitem{ozfatura2019speeding}
E.~Ozfatura, D.~G{\"u}nd{\"u}z, and S.~Ulukus, ``Speeding up distributed
  gradient descent by utilizing non-persistent stragglers,'' in \emph{2019 IEEE
  International Symposium on Information Theory (ISIT)}.\hskip 1em plus 0.5em
  minus 0.4em\relax IEEE, 2019, pp. 2729--2733.

\bibitem{maity2019robust}
R.~K. Maity, A.~S. Rawa, and A.~Mazumdar, ``Robust gradient descent via moment
  encoding and ldpc codes,'' in \emph{2019 IEEE International Symposium on
  Information Theory (ISIT)}.\hskip 1em plus 0.5em minus 0.4em\relax IEEE,
  2019, pp. 2734--2738.

\bibitem{yu2020straggler}
Q.~Yu, M.~A. Maddah-Ali, and A.~S. Avestimehr, ``Straggler mitigation in
  distributed matrix multiplication: Fundamental limits and optimal coding,''
  \emph{IEEE Transactions on Information Theory}, vol.~66, no.~3, pp.
  1920--1933, 2020.

\bibitem{ho2013more}
Q.~Ho, J.~Cipar, H.~Cui, S.~Lee, J.~K. Kim, P.~B. Gibbons, G.~A. Gibson,
  G.~Ganger, and E.~P. Xing, ``More effective distributed ml via a stale
  synchronous parallel parameter server,'' in \emph{Advances in neural
  information processing systems}, 2013, pp. 1223--1231.

\bibitem{dean2008mapreduce}
J.~Dean and S.~Ghemawat, ``Mapreduce: simplified data processing on large
  clusters,'' \emph{Communications of the ACM}, vol.~51, no.~1, pp. 107--113,
  2008.

\bibitem{zaharia2008improving}
M.~Zaharia, A.~Konwinski, A.~D. Joseph, R.~H. Katz, and I.~Stoica, ``Improving
  mapreduce performance in heterogeneous environments.'' in \emph{Osdi},
  vol.~8, no.~4, 2008, p.~7.

\bibitem{lee2017speeding}
K.~Lee, M.~Lam, R.~Pedarsani, D.~Papailiopoulos, and K.~Ramchandran, ``Speeding
  up distributed machine learning using codes,'' \emph{IEEE Transactions on
  Information Theory}, vol.~64, no.~3, pp. 1514--1529, 2017.

\bibitem{tandon2017gradient}
R.~Tandon, Q.~Lei, A.~G. Dimakis, and N.~Karampatziakis, ``Gradient coding:
  Avoiding stragglers in distributed learning,'' in \emph{International
  Conference on Machine Learning}, 2017, pp. 3368--3376.

\bibitem{neighbors1984draco}
J.~M. Neighbors, ``The draco approach to constructing software from reusable
  components,'' \emph{IEEE Transactions on Software Engineering}, no.~5, pp.
  564--574, 1984.

\bibitem{halbawi2018improving}
W.~Halbawi, N.~Azizan, F.~Salehi, and B.~Hassibi, ``Improving distributed
  gradient descent using reed-solomon codes,'' in \emph{2018 IEEE International
  Symposium on Information Theory (ISIT)}.\hskip 1em plus 0.5em minus
  0.4em\relax IEEE, 2018, pp. 2027--2031.

\bibitem{dutta2018unified}
S.~Dutta, Z.~Bai, H.~Jeong, T.~M. Low, and P.~Grover, ``A unified coded deep
  neural network training strategy based on generalized polydot codes,'' in
  \emph{2018 IEEE International Symposium on Information Theory (ISIT)}.\hskip
  1em plus 0.5em minus 0.4em\relax IEEE, 2018, pp. 1585--1589.

\bibitem{royston1982algorithm}
J.~Royston, ``Algorithm as 177: Expected normal order statistics (exact and
  approximate),'' \emph{Journal of the royal statistical society. Series C
  (Applied statistics)}, vol.~31, no.~2, pp. 161--165, 1982.

\end{thebibliography}
\end{document}